\documentclass[%
 reprint,
 amsmath,amssymb,
 aps,
]{revtex4-2}
\usepackage{amsmath}
\usepackage{graphicx}
\usepackage{dcolumn}
\usepackage{url}
\usepackage{xcolor}
\usepackage{hyperref}

\newcommand{\blue}[1]{\textcolor{blue}{#1}}
\usepackage{hyperref}
\usepackage{appendix}
\usepackage{amsthm}
\usepackage[ruled,linesnumbered]{algorithm2e}
\hypersetup{
    colorlinks=true,  
    linkcolor=blue,   
    filecolor=blue,   
    urlcolor=blue,    
    pdftitle={Correlated error can impro},  
    pdfauthor={Author}, 
}
\newtheorem{theorem}{Theorem}[section]  
\newtheorem{lemma}[theorem]{Lemma}      
\begin{document}

\preprint{APS/123-QED}
\title{Symmetry in Multi-Qubit Correlated Noise Errors Enhances Surface Code Thresholds}

\author{SiYing Wang}
\author{Yue Yan}
\author{ZhiXin Xia}
\author{Xiang-Bin Wang}
\email{ xbwang@mail.tsinghua.edu.cn}
\affiliation{ State Key Laboratory of Low Dimensional Quantum Physics, Department of Physics, \\ Tsinghua University, Beijing 100084, China}

\date{\today}

\begin{abstract}
Surface codes are promising for practical quantum error correction due to their high threshold and experimental feasibility. However, their performance under realistic noise conditions, particularly those involving correlated errors, requires further investigation. In this study, we investigate the impact of correlated errors on the error threshold. In particular, we focus on several distinct types of correlated errors that could potentially arise from next-nearest-neighbor (NNN) coupling in quantum systems. We present the analytical threshold of the surface code under these types of correlated noise, and find that errors correlated along straight lines possess a type of crucial symmetry, resulting in higher thresholds compared to other types of correlated errors. This deepens our insight into the threshold of surface code and hence facilitates a more robust design of quantum circuits with a higher noise threshold.
\end{abstract}

\maketitle


\section{\label{sec:intro}introduction}
High-performance quantum error correction codes are essential for realizing practical quantum computing with noise. Among these, the surface code is a promising candidate that has been extensively studied\cite{quantum_codes1,quantum_codes2,quantum_codes3,quantum_codes4,quantum_codes5}. It is characterized by a high error threshold and utilizes gates performed between nearest-neighbor qubits arranged in a two-dimensional grid\cite{Surface_code1,Surface_code_threshold1,Surface_code_threshold3,Surface_code_threshold4,xzzx_2021}. This makes the surface code particularly suitable for experimental implementations, as evidenced by recent demonstrations\cite{quantum_codes1,quantum_codes2,quantum_codes4,Surface_code_threshold5}.
\par
The threshold of the surface code is related to the system noise,  therefore, it is crucial to evaluate the performance of the surface code under realistic and effective error models\cite{correlatedthreshold_thresholdprl,correlatedthreshold_spatially_2024,Surface_code_threshold1,correlatedthreshold_breakdown_2014}. In real experimental setups, there are often correlated errors such as unwanted crosstalk and coupling to a common Bose bath\cite{experiment_realizing_2022,experiment_correlated_2021,experiment_mid-circuit_2022,experiment1,experiment2}. These correlations can significantly decrease the threshold of the surface code and many codes break down under correlated noise \cite{correlated_fowler2014quantifying,correlated_quantum_2019}. \par
The performance of surface codes under correlated errors have been explored in the past, however, most research has been limited to numerical study\cite{correlatedthreshold_impact_2021,correlatedthreshold_spatially_2024,correlated_fowler2014quantifying,correlated_quantum_2019}. The influence of correlated error on the threshold have been also investigated\cite{correlatedthreshold_thresholdprl}, but it is limited to the specific type of correlated error due to Bose baths. \par
In this work, we present the thresholds of surface codes under different types of correlated error models. We first show that a class of error models with correlations along straight lines possesses a crucial symmetry that can enhance the threshold of surface codes. In particular, this symmetry enables the surface code to have a higher threshold for this type of correlated error compared to all other types of correlated errors studied in this work. This fact deepens our insight to threshold of surface code and hence facilitates to more robust design of quantum circuits with higher noise threshold through compressing those types of correlated errors with lower threshold.\par
Our paper is organized as follows. We begin by  introducing the planar code and the correlated error model in Sec.\ref{Sec:Noise}.  In Sec.\ref{Sec:Threshold}, we first demonstrated that correlated errors meeting certain conditions might increase the threshold of the surface code. We then use a syndrome-based equivalence approach to calculate thresholds of the other types of surface code. In Sec.\ref{Sec:Num}, we used Stim to simulate the impact of these correlated errors on the surface code threshold at code capacity and the circuit level. Finally, we summarize our key findings and discuss possible future work in Sec.\ref{Sec:conclusion}.\par
\section{The planar code and Noise model\label{Sec:Noise}}
\subsection{Planar code}
Planar codes are a class of topological quantum error correcting codes. They are defined on a $d \times d$  two-dimensional lattice with open boundaries, where $d$ denotes the distance of surface code, $n$ denotes the number of physical qubits which are associated with edges on the lattice. Following conventional notations, we define the stabilizer generators as follows: $A_v=\prod_{e \in v} X_e$ represents the product of $X$ operators on edges surrounding vertices, while $B_p=\prod_{e \in p} Z_e$ denotes the product of $Z$ operators on edges enclosing plaquettes. Consequently, the stabilizer group is defined as $\mathcal{G}=\left\langle A_v, B_p\right\rangle$. These logical operators $\bar{X}$ and $\bar{Z}$ satisfy $\bar{X}, \bar{Z} \in \mathcal{C}(\mathcal{G}) \backslash \mathcal{G}$ and $\overline{X Z}=-\overline{Z X}$, where $\mathcal{C}(\mathcal{G})=\{f \in \mathcal{P}: f g=g f  ,\forall g \in \mathcal{G}\}$ denotes the centralizer of $\mathcal{G}$, and $\mathcal{P}$ represents the group of $n$-qubit Pauli operators. \cite{Surface_code_threshold1,kitaev_surfacecode}. 

When an error $E$ occurs on a Surface code, it anticommutes with certain stabilizer generators $g$, resulting in the manifestation of a syndrome. We can formally define the syndrome $s$ as a binary vector where $\{s_i\in s:\forall g_i\in\mathcal{G},  E^{\dagger}g_iE=s_ig_i\} $

The decoding process aims to identify a recovery operator $R$ that produces the same syndrome as the error $E$.  It's important to note that the recovery operator $R$ need not be identical to $E$,  all recovery operators of the form $RE\in \mathcal{C}(\mathcal{G})$  yield identical syndromes. Within this framework, the decoding process leads to a logical error and consequently fails when $RE \in \mathcal{C}(\mathcal{G}) \backslash \mathcal{G}$. Conversely, error correction succeeds when $RE \in \mathcal{G}$. 
\subsection{Noise model \label{Noise model}}
We focuses on correlated errors between data qubits in surface codes. Since $X$ errors and $Z$ errors are corrected independently in surface codes, we focus our analysis on scenarios with only $Z$ errors. Of course, those $X$ errors can be treated similarly. For clarity, we use the following error model for $Z$ errors.\par
\begin{figure}
    \centering
    \includegraphics[width=1\linewidth]{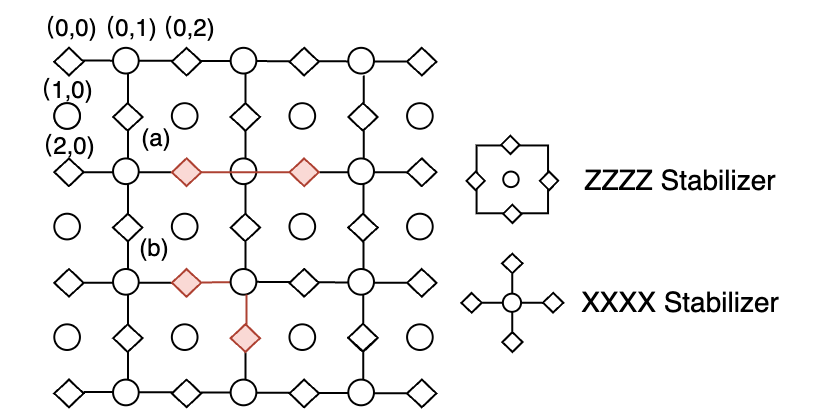}
    \caption{The planar code and two types of correlated noise models. The circles denote ancilla qubits, while the diamonds represent data qubits. The indices of some qubits are shown in the figure. (a): Example for 2-qubit correlated error of type-1. The two pink diamonds located on the line represent two $Z$ errors simultaneously occurring on a pair of qubits. (b): Example for type-2 correlated error. The two adjacent pink diamonds represent two $Z$ errors simultaneously occurring on a pair of qubits.}
    \label{fig:surface code and error model}
\end{figure}

Define two sets $Q_1,Q_2$:
\begin{align*}
    &Q_1=\{q_1(i,j),q_2(i,j)|0<i<d-k+1,0<j<d-k+1\}\\
    &\begin{cases}&q_1(i,j)=\{(2i,2j),(2i,2j+2),\cdots,(2i,2i+2k-2)\} \\&q_2(i,j)=\{(2i+1,2j+1),\cdots,(2i+2k-1,2j+1)\}
    \end{cases}\\
    &Q_2=\{q_3(i,j),q_4(i,j)|0<i<d-1,1<j<d-1\}\\
    &\begin{cases}&q_3(i,j)=\{(2i,2j),(2i+1,2j+1)\}\\&q_4(i,j)=\{(2i,2j),(2i+1,2j-1)\}\end{cases}
\end{align*}
where $(i,j)$ is the indice  of qubit in a two-dimensional. Some of these indices are shown in Fig.\ref{fig:surface code and error model}. Here, $i$ and $j$ are either both odd or both even, corresponding to the coordinates of data qubits. As shown in Fig.\ref{fig:surface code and error model}, $q_1$ and $q_2$ represent sets of k data qubits on a straight line, while $q_3$ and $q_4$ represent pairs of neighboring qubits. Then we can use these sets to describe the error model, which can be defined as follows:
\begin{itemize}
    \item Type-1 correlated error: Each qubit corresponding to an index in $q_1$ or $q_2$ has a probability $p_1$ of experiencing a phase flip error, which can be described as:
    $$
    \mathcal{E}(\rho)=\left(1-p_1\right) \rho+p_1 Z_1Z_2\cdots Z_n\rho Z^{\dagger}_kZ^{\dagger}_{k-1}\cdots Z^{\dagger}_1
    $$ 
\item Type-2 correlated error: Each qubit corresponding to an index in $q_3$ and $q_4$ has a probability $p_2$ of experiencing a phase flip error, which can be described as:
    $$
\mathcal{E}(\rho)=\left(1-p_2\right) \rho+p_2 Z_1Z_2\rho Z^{\dagger}_2 Z^{\dagger}_1
$$ 
\end{itemize}\par
These types of data-data qubits correlated noise frequently appears in quantum computing\cite{correlated_2023learning,correlated_tiurev2023correcting}, for example, in superconducting quantum architectures of surface code, the Next-Nearest-Neighbor (NNN) coupling can cause correlation between data-data qubits and lead to correlated errors occurring simultaneously on two data qubits.\cite{NNNcrosstalk_marxer2023long,NNNcrosstalk_marxer2023long}. Obviously, understanding the impact of different types of correlated errors is essential for improving performance of surface code.
\section{the Analytic Thresholds of surface code under correlated error\label{Sec:Threshold}}
\subsection{Symmetry of noise model\label{Sec:Symmetry}}
\begin{figure}
    \centering
    \includegraphics[width=0.7\linewidth]{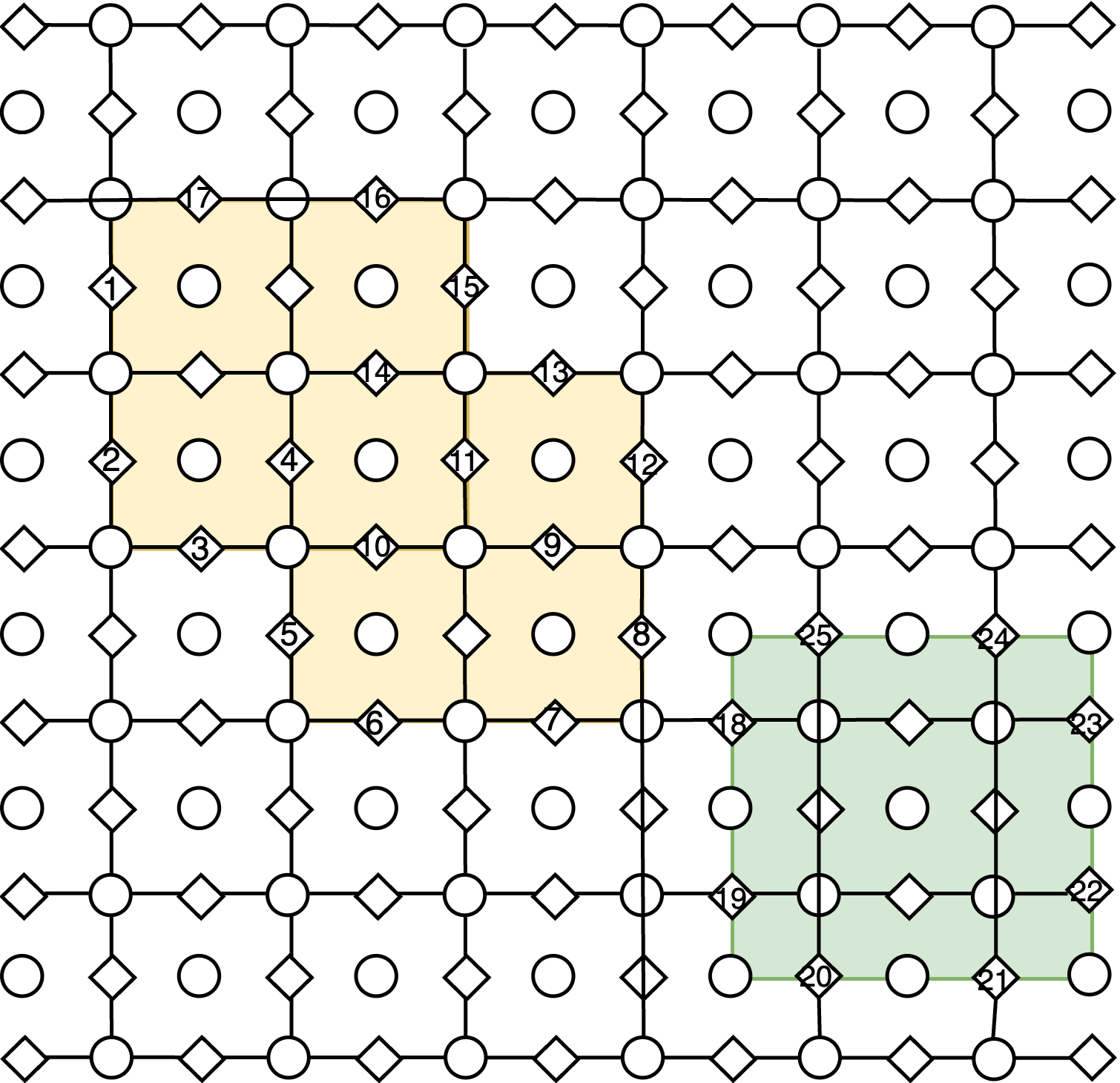}
    \caption{Examples of elements in set $S_{sys}$. For Z-error scenarios, elements in the yellow region such as $Z_1Z_2Z_3Z_{10}Z_{11}Z_{15}Z_{16}Z_{17}$ and $Z_4Z_5Z_6Z_{7}Z_{8}Z_{12}Z_{13}Z_{14}$ belong to $S_{sys}$. For X-error scenarios, elements in the green region such as $X_{18}X_{19}X_{20}X_{21}X_{22}X_{23}X_{24}X_{25}$ are included in $S_{sys}$.}
    \label{fig:symmmetry}
\end{figure}

In this chapter, we will demonstrate that the threshold of surface codes increases when errors possess certain symmetries and use type-1 correlated errors as an example. Denote the group composed of all errors as \(\mathcal{E} = \{E_1, E_2, \ldots\}\), and the elements in \(\mathcal{E}\) all commute with each other. When errors in $\mathcal{E}$ are i.i.d, \(\mathcal{C}(\mathcal{G}) \cap \mathcal{E} = \mathcal{C}(\mathcal{G})\). In this case, we need to seek recovery within the coset of \(\mathcal{C}(\mathcal{G})\). However, if errors in \(\mathcal{E}\) is not i.i.d, \(\mathcal{C}(\mathcal{G}) \cap \mathcal{E} \neq \mathcal{C}(\mathcal{G})\), and in this case we denote 
\begin{equation}\label{symmetry}
    S_{sys}=\mathcal{C}(\mathcal{G}) \cap \mathcal{E}
\end{equation}
Here, evidently, $S_{sys}$ is a subgroup of $\mathcal{C}(\mathcal{G})$. Under this situation, we only need to find the recovery within cosets of \(\mathcal{S}_{sys}\). It is straightforward to prove that when \(i \in \mathcal{S}_{sys}\), we have \(E^{\dagger}iE = i\). The definition of \(\mathcal{S}_{sys}\) here generalizes Ref.\cite{decoder1,decoder2} where symmetries are defined for $E^{\dagger}A_vE = A_v, E^{\dagger}B_pE =B_p$. Therefore, we establish the important formula for succeeding probability of our error correction.
\begin{equation}
    P_{success}=\frac{P(RE\in S_{sys}\cap\mathcal{G})}{P(RE\in S_{sys}\cap\mathcal{C}(\mathcal{G}))}
\end{equation}
By this formula, it is easy to see that under conditions where most elements in $S_{sys}$ belong to $\mathcal{G}$. The ratio \(\frac{P(RE \in S_{sys} \cap \mathcal{G})}{P(RE \in S_{sys} \cap \mathcal{C}(\mathcal{G}))}\) has the potential to be greater than \(\frac{P(RE \in \mathcal{G})}{P(RE \in \mathcal{C}(\mathcal{G}))}\).  We will demonstrate that type-1 correlated Z or X errors along straight lines exhibit a symmetry that can increase $ P_{success}$. 
The elements of $S_{sym}$ are then constructed by taking the product of stabilizers within square regions of side length $k$. The case where $k = 2$ is shown in Fig.\ref{fig:symmmetry}. 
Now we show how $P_{success}$ evolves under the symmetry of Eq.\eqref{symmetry}. Evidently, $S_{sym}$ excludes logical operators \blue{when }
\begin{equation}\label{correlation-lenth}
    d/j\not\in \mathbb{Z}.
\end{equation}
In this case, $P_{success}=1$ and the threshold of surface code is 
\begin{equation}\label{threshold3}
    p_{th}=1.
\end{equation}
This offers a promising outlook, suggesting that correlated noise of length $j$ is beneficial in error correction. Also, this result infers that when $p_1 \neq 0$, as long as $\frac{p_j}{p_1}$ is sufficiently large, we can achieve a lower logical error rate than that in the i.i.d. noise channel.
\subsection{A Syndrome-Based Equivalence Approach\label{Sec:approach}}
For the noise model in Sec.\ref{Sec:Noise}, it is difficult to write an analytical expression for the probability of an error chain. Therefore, we use a syndrome-based equivalence approach. We start with notations for lemma.\ref{lemma1} below.\par
$\mathcal{E}$: an error group that can be decomposed into the direct product of two subgroups $\mathcal{E}_1$ and $\mathcal{E}_2$, that is, $\mathcal{E}=\mathcal{E}_1 \times \mathcal{E}_2$; $p_{th}$: the threshold of the surface code under error $\mathcal{E}$; $p_{th1}(p_{th2})$: the threshold under error $\mathcal{E}_1(\mathcal{E}_2)$.
\begin{lemma}\label{lemma1}
     $p_{th}=min\{p_{th1},p_{th2}\}$ provided that $E_1\mathcal{C}(\mathcal{G})\notin \mathcal{E}_2$ for any $E_1 \in \mathcal{E}_1$.
\end{lemma}
Proof: given $E_1\mathcal{C}(\mathcal{G})\notin \mathcal{E}_2$ for any $E_1 \in \mathcal{E}_1$, we have $\mathcal{S}_1\cap\mathcal{S}_2=\emptyset$, where $\mathcal{S}_1(\mathcal{S}_2)$ is the set that contains syndromes produced by $\mathcal{E}_1(\mathcal{E}_2)$. Consequently, we can identify which error model $(\mathcal{E}_1,\mathcal{E}_2)$ has caused the error based on the observed syndrome. Given this, we can actually correct errors independently due to $\mathcal{E}_1$ and $\mathcal{E}_2$. This completes the proof.\par
\begin{figure}
    \centering
    \includegraphics[width=1\linewidth]{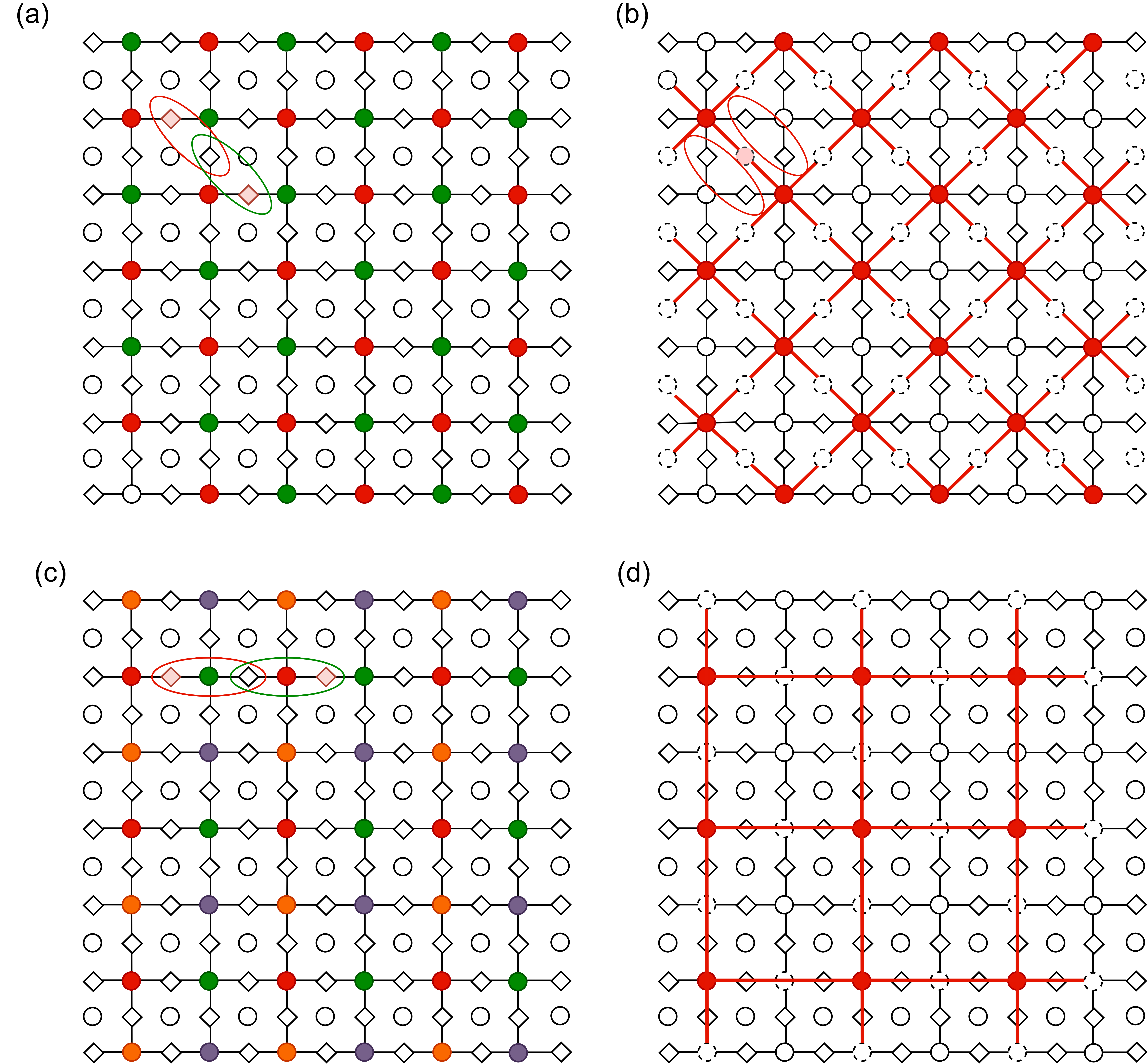}
    \caption{Decomposition of correlated error models. (a) Type-2 correlated errors can be decomposed into two error models: one causing syndromes at green nodes and the other causing syndromes at red nodes. (b) Equivalent surface code after map $\mathcal{M}$. When errors simultaneously occur on the two qubit pairs encircled by pink ellipses, it corresponds to an error occurring on the virtual qubit represented by the pink dashed circle. (c) Type-1 correlated errors are decomposed into four error models, each producing syndromes at different colored nodes: orange, purple, red, and green respectively. (d) Equivalent surface code after map $\mathcal{M}$. Dashed circles represent virtual qubits, and red points are ancilla qubits.}
    \label{divide error}
\end{figure}
\begin{figure}
    \centering
    \includegraphics[width=1\linewidth]{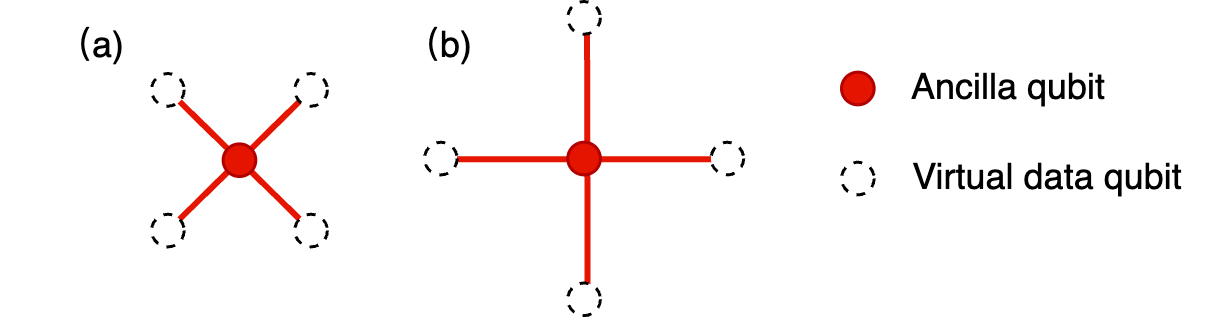}
    \caption{Equivalent stabilizers after mapping correlated errors to i.i.d. errors. The stabilizer is an operator $ZZZZ$ acting on four virtual qubits. (a): stabilizer resulting from mapping type-2 correlated errors. (b): stabilizer resulting from mapping type-1 correlated errors.}
    \label{Map}
\end{figure}

Since all the noise models in Sec.\ref{Noise model} satisfy the conditions of lemma.\ref{lemma1}, it is applicable to the error thresholds of these models.
We first analyze type-2 errors, which can be decomposed into the direct product of two subgroups $\mathcal{E}_1$ and
$\mathcal{E}_2$. As shown in Fig.\ref{divide error}(a), elements $E \in \mathcal{E}_1$ will cause syndromes marked by red dots, and elements $E \in \mathcal{E}_2$ will cause syndromes marked by green dots. For example,  the error circled by the pink ellipse belongs to $\mathcal{E}_1$ and the error circled by the green ellipse belongs to $\mathcal{E}_2$. The red syndrome and green syndrome do not overlap, so we can decode errors produced by $\mathcal{E}_1$ using the red syndrome, and decode errors produced by $\mathcal{E}_2$ using the green syndrome separately. In this case, the overlapped correlated errors described in Sec.\ref{Noise model} are equivalent to i.i.d. errors.
Because of the obvious symmetry in Fig.\ref{divide error}(a), the threshold of $\mathcal{E}_2$ is the same as $\mathcal{E}_1$, that is $p_{th1}=p_{th2}$. So we only take $\mathcal{E}_1$ as an example for threshold analysis. We map the correlated error of a pair of nearest data-data qubits to a virtual qubit (dashed circle), as shown in Fig.\ref{divide error}(b). For convenience of presentation, we use the notation $\mathcal{M}$ to represent this map. Each virtual qubit induces syndromes at two adjacent red nodes. Corresponding to the map  $\mathcal{M}$, the stabilizers of surface code is transformed accordingly, as is shown in Fig.\ref{Map}, which is an operator $X_1X_2X_3X_4$ acting on four virtual qubits. \par
The new surface code after map  $\mathcal{M}$ above in Fig.\ref{divide error}(b) differs from the original surface code shown in Fig.\ref{divide error}(a) only in size and boundary conditions. Since the threshold is defined for an infinite size, the boundary conditions and size do not affect the threshold of the surface code. Therefore, the threshold of the new surface code in Fig.\ref{divide error}(b) is the same as that of the original surface code in Fig.\ref{divide error}(a). Under i.i.d. model, the noise in the surface code can be obtained through statistical mapping\cite{Surface_code1,Surface_code_threshold1}. In particular, under  i.i.d $Z$ error, the exact threshold value for the surface code is $p_{th}=10.9\%$\cite{Surface_code1,surface_code_nishimori}. As shown in Fig.\ref{divide error}(b), an error occurring in either of the two encircled pairs of nearest data qubits will induce an error in the same virtual qubit. Moreover, when errors occur simultaneously in both of the encircled pairs, they generate a stabilizer, which does not change the state of the surface code. In this scenario, there is no error in the virtual data qubit. Therefore, we conclude that the error probability for the virtual qubits is $2p_2(1-p_2)$. And hence, the threshold of the surface code under type-2 correlated noise is $2p_{th}(1-p_{th})=10.9$, which means
\begin{align}\label{threshold1}
    p_{th}=5.8\%.
\end{align}

Now we analyze the threshold of another form of type-1  correlated errors, $d/j\in \mathbb{Z}$. Similar to type-2 correlated errors above, here we can decompose the error model in \ref{Noise model} into four independent error models in Fig.\ref{divide error}(c), and they can be independently decoded given syndrome nodes of different colors. We take the error model that generates red syndromes for threshold analysis for example, as illustrated in Fig.\ref{divide error}(d). The correlated 2-qubit noise can be mapped to i.i.d. noise. The stabilizers of the surface code after this map are shown in Fig.\ref{Map}. The error probability for each virtual qubit is equivalent to the probability of a correlated error occurring on a pair of data qubits $p_1$. Therefore, for type-1 correlated errors where $d/j \in \mathbb{Z}$, the threshold is identical to that of i.i.d. noise: 
\begin{align}\label{threshold2}
    p_{th} = 10.9\%
\end{align}

\begin{figure}
    \centering
    \includegraphics[width=1\linewidth]{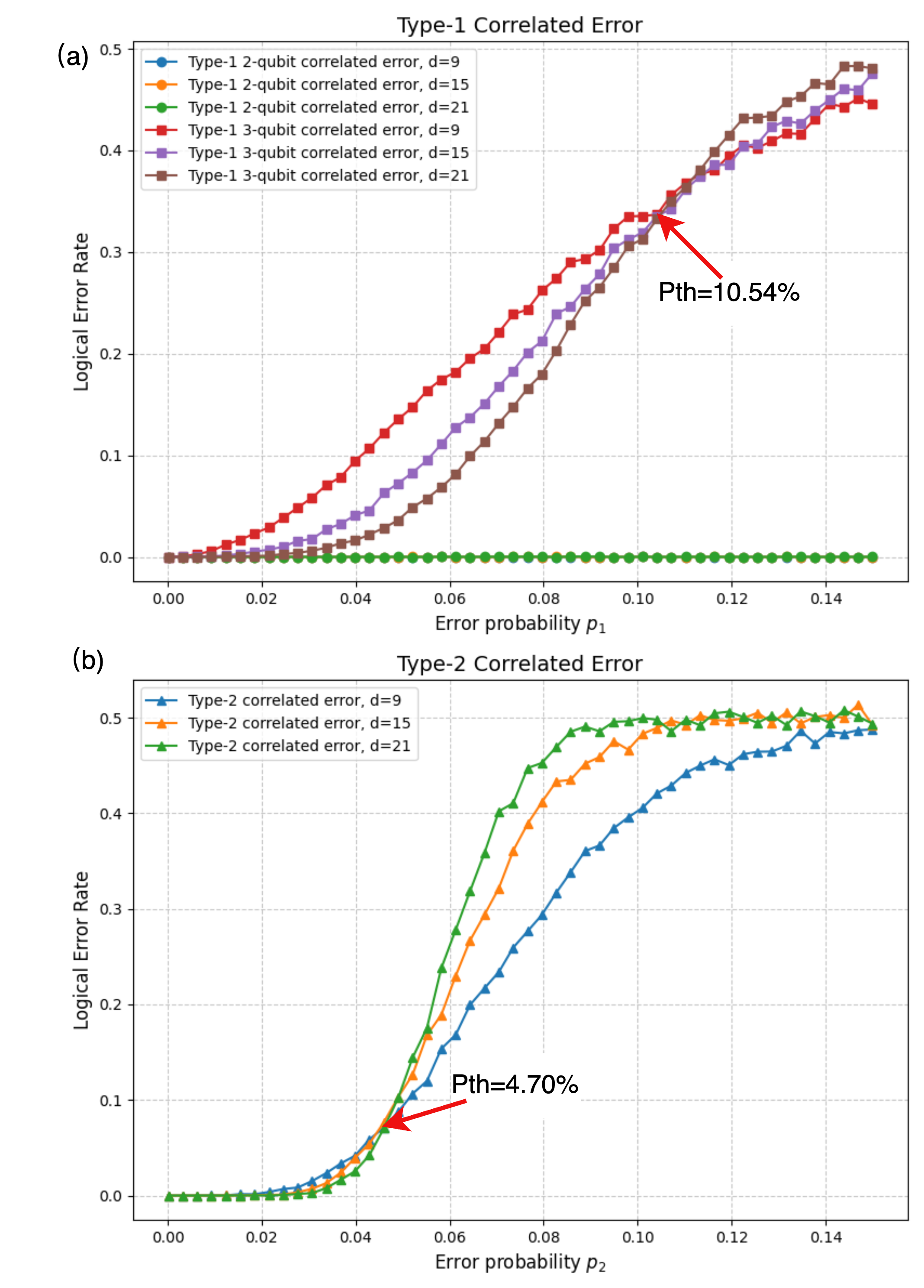}
    \caption{Decoding results of Pymatching 2.0 in logical error rates for two types of correlated errors, with $d = 9, 15, 21$. (a): 2 qubit correlated errors and 3 qubit correlated errors of type-1 as a function of their occurring probability $p_1$. The 2 qubit correlated errors of type-1 are perfectly corrected without producing logical errors, while the threshold for 3 qubit correlated errors of type-1 is 10.54\%. (b): type-2 correlated errors as a function of its occurring probability $p_2$. The threshold is 4.7\%.}
    \label{correlated threshold}
\end{figure}
\begin{figure*}
    \centering
    \includegraphics[width=1\linewidth]{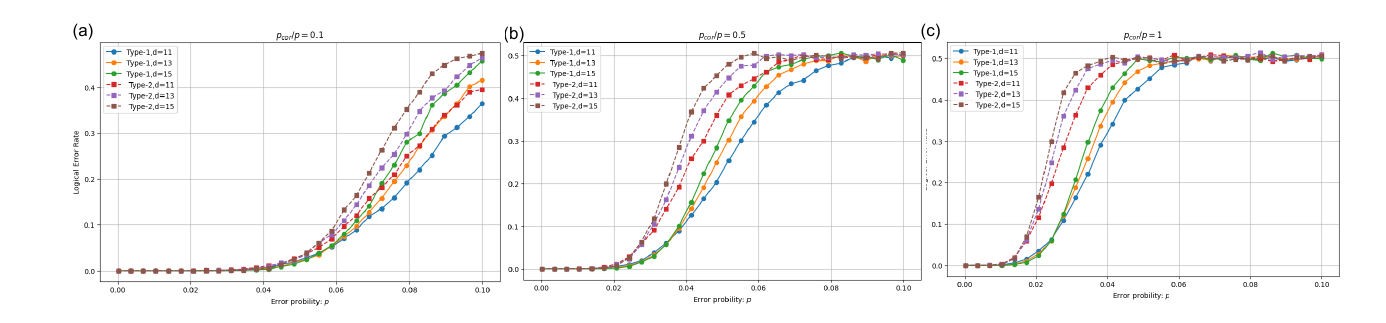}
    \caption{Logical error rate of the $d=11,13,15$ surface codes as a function of error probability $p$. The solid lines represent the simulation results for the scenario where circuit level noise and 2-qubit correlated errors of type-1 coexist. The dashed lines represent the simulation results for the scenario where circuit level noise and type-2 correlated errors coexist. (a): $p_{cor}/p=0.1$. (b): $p_{cor}/p=0.5$. (c): $p_{cor}/p=1$.}
    \label{fig:circuit_threshold}
\end{figure*}
\section{code performance under correlated error\label{Sec:Num}}
In this section, we use Stim to simulate circuit-level errors in the surface code, and employ PyMatching 2.0 for decoding \cite{gidney2021stim,pymatching2,higgott2022pymatching}. PyMatching 2.0 generates a detector graph based on the input error model. It reassigns weights to the relevant edges for correlated errors, thus possessing the capability to handle correlated errors that produce two syndromes.\par
As a test of our analysis, we simulated the logical error rate of surface codes  under correlated noise models in Sec.\ref{Noise model}. As presented in Fig.\ref{correlated threshold}, the decoding results are rather close or equal to our upper bounds in Eqs.(\ref{threshold1},\ref{threshold2},\ref{threshold3}). This validates our results in Eqs.(\ref{threshold1},\ref{threshold2},\ref{threshold3}) and also shows the excellent performance of PyMatching 2.0. In particular, three curves in Fig.\ref{correlated threshold} perfectly match the predictions in Eq.\ref{threshold3}. In Fig.\ref{correlated threshold}(a), the 2-qubit correlated errors of type-1 satisfy $d/j \notin Z$, and therefore perfectly correctable, these clearly in accordance with our results in Eq.(\ref{threshold3}). The result indicate that type-1 2-qubit correlated errors can be perfectly corrected in $d/j \notin Z$ during error correction, suggesting that when this type of error is predominant, the logical error rate of the surface code will be improved.  3-qubit correlated errors of type-1 satisfy $d/j\in Z$, with a numerically simulated threshold of $10.54\%$, close to the i.i.d. threshold of $10.9\%$. In Fig.\ref{correlated threshold}(b), the error correction threshold for type-2 correlated errors is $4.7\%$, close to our predicted upper bound of $5.8\%$. 
\par
Now we consider a more general case where four types of errors below coexist. Here we use the noise mode of Ref.\cite{circuit_2023improved,circuit_chamberland2022universal} to describe the circuit-level noise:
\begin{enumerate}
    \item Each two-qubit gate is followed by a two-qubit Pauli channel:
    $$
\mathcal{E}(\rho)=\left(1-p\right) \rho+\frac{p}{15} \sum_{i, j}\left[\sigma^i \otimes \sigma^j\right] \rho\left[\sigma^i \otimes \sigma^j\right]
$$
where $i, j \in\{I, X, Y, Z\}$ and $(i, j) \neq(I, I)$.
    \item Each single-qubit gate location or single-qubit idle location of the same duration (a single time step) is followed by a Z error with probability $p$.
    \item Each single-qubit X-basis measurement is flipped with probability $p$.
    \item Each two-qubit gate is followed by type-1 2 qubit correlated or type-2 correlated error with probability $p_{cor}$
\end{enumerate}
As shown in Fig. \ref{fig:circuit_threshold}, we simulated the logical error rates of the surface code for different values of $p_{cor}/p$. As a result, when both data-data correlated errors and circuit-level noise coexist, the threshold for 2-qubit correlated errors of type-1 are higher than in the case where type-2 correlated errors exist. This shows that the performance of surface codes will be significantly improved if the occurring probability of type-2 correlated errors is higher than that of type-1 correlated errors.

Taking the superconducting quantum computing architecture as an example, we can suppress NNN coupling by detuning frequency  between qubits\cite{detuning_kelly2014optimal,detuning_zhao2022quantum}. However, qubit frequencies can only be chosen within a specific range\cite{detuning_zhao2022quantum}. Given this constraint, it's challenging to achieve sufficient detuning between qubits to eliminate all correlated errors. Consequently, the frequency distribution of qubits must be strategically designed to modulate the probabilities of different types of correlated errors. We can choose to set a larger dunning only between neighboring qubits and a no detuning between next neighboring qubits, i.e. no detuning for qubits along a straight line. Given such a choice, the type-2 correlated errors would be suppressed and the type-1 correlated errors would be increased. This shows an application example for our results to improve the error correction performance of the surface code.
\section{Conclusion\label{Sec:conclusion}}
In summary, through analytical study, we obtain explicit results for thresholds of surface codes under different types of correlated errors described in Sec.\ref{Noise model}. We show that the correlated errors along a linear configuration possess a symmetry that enables a higher threshold of the surface code. Furthermore, we study the effects of different types of correlated errors at the circuit level on the threshold and logical error rate of surface codes. Our results conclude that these symmetrical correlated errors lead to higher thresholds and lower logical error rates for surface codes. These results are extensivly tested numerically. Our results show that type-2 correlated errors significantly reduce the threshold of surface codes. This means that the performance of surface codes in practical quantum computing will be improved if type-2 correlated errors is reduced.

\begin{acknowledgements}
We acknowledge the financial support in part by National Natural Science Foundation of China grant No.12174215 and No.12374473, and Innovation Program for Quantum Science and Technology No.2021ZD0300705. This study is also supported by the Taishan Scholars Program. 
\end{acknowledgements}
\bibliographystyle{unsrt}
\bibliography{ref.bib}

\begin{thebibliography}{10}

\bibitem{quantum_codes1}
Google~Quantum AI.
\newblock Suppressing quantum errors by scaling a surface code logical qubit.
\newblock {\em Nature}, 614(7949):676--681, February 2023.

\bibitem{quantum_codes2}
Sebastian Krinner, Nathan Lacroix, Ants Remm, Agustin Di~Paolo, Elie Genois, Catherine Leroux, Christoph Hellings, Stefania Lazar, Francois Swiadek, Johannes Herrmann, et~al.
\newblock Realizing repeated quantum error correction in a distance-three surface code.
\newblock {\em Nature}, 605(7911):669--674, 2022.

\bibitem{quantum_codes3}
Andr{\'a}s Gyenis, Pranav~S Mundada, Agustin Di~Paolo, Thomas~M Hazard, Xinyuan You, David~I Schuster, Jens Koch, Alexandre Blais, and Andrew~A Houck.
\newblock Experimental realization of a protected superconducting circuit derived from the 0--$\pi$ qubit.
\newblock {\em PRX Quantum}, 2(1):010339, 2021.

\bibitem{quantum_codes4}
Sebastian Krinner, Nathan Lacroix, Ants Remm, Agustin Di~Paolo, Elie Genois, Catherine Leroux, Christoph Hellings, Stefania Lazar, Francois Swiadek, Johannes Herrmann, Graham~J. Norris, Christian~Kraglund Andersen, Markus Müller, Alexandre Blais, Christopher Eichler, and Andreas Wallraff.
\newblock Realizing {Repeated} {Quantum} {Error} {Correction} in a {Distance}-{Three} {Surface} {Code}.
\newblock {\em Nature}, 605(7911), May 2022.
\newblock arXiv:2112.03708 [cond-mat, physics:quant-ph].

\bibitem{quantum_codes5}
{Google Quantum AI}.
\newblock Exponential suppression of bit or phase errors with cyclic error correction.
\newblock {\em Nature}, 595(7867):383--387, July 2021.

\bibitem{Surface_code1}
Eric Dennis, Alexei Kitaev, Andrew Landahl, and John Preskill.
\newblock Topological quantum memory.
\newblock {\em Journal of Mathematical Physics}, 43(9):4452--4505, September 2002.
\newblock arXiv:quant-ph/0110143.

\bibitem{Surface_code_threshold1}
H.~Bombin, Ruben~S. Andrist, Masayuki Ohzeki, Helmut~G. Katzgraber, and M.~A. Martin-Delgado.
\newblock Strong {Resilience} of {Topological} {Codes} to {Depolarization}.
\newblock {\em Physical Review X}, 2(2):021004, April 2012.

\bibitem{Surface_code_threshold3}
Austin~G. Fowler.
\newblock Analytic asymptotic performance of topological codes.
\newblock {\em Physical Review A}, 87(4):040301, April 2013.
\newblock arXiv:1208.1334 [quant-ph].

\bibitem{Surface_code_threshold4}
Austin~G. Fowler.
\newblock Proof of finite surface code threshold for matching.
\newblock {\em Physical Review Letters}, 109(18):180502, November 2012.
\newblock arXiv:1206.0800 [quant-ph].

\bibitem{xzzx_2021}
J.~Pablo Bonilla~Ataides, David~K. Tuckett, Stephen~D. Bartlett, Steven~T. Flammia, and Benjamin~J. Brown.
\newblock The {XZZX} surface code.
\newblock {\em Nature Communications}, 12(1):2172, April 2021.

\bibitem{Surface_code_threshold5}
Rajeev Acharya, Laleh Aghababaie-Beni, and Aleiner et~al.
\newblock Quantum error correction below the surface code threshold, August 2024.
\newblock arXiv:2408.13687.

\bibitem{correlatedthreshold_thresholdprl}
E.~Novais and Eduardo~R. Mucciolo.
\newblock Surface {Code} {Threshold} in the {Presence} of {Correlated} {Errors}.
\newblock {\em Physical Review Letters}, 110(1):010502, January 2013.

\bibitem{correlatedthreshold_spatially_2024}
Ji~Zou, Stefano Bosco, and Daniel Loss.
\newblock Spatially correlated classical and quantum noise in driven qubits.
\newblock {\em npj Quantum Information}, 10(1):46, April 2024.

\bibitem{correlatedthreshold_breakdown_2014}
Adrian Hutter and Daniel Loss.
\newblock Breakdown of {Surface} {Code} {Error} {Correction} {Due} to {Coupling} to a {Bosonic} {Bath}.
\newblock {\em Physical Review A}, 89(4):042334, April 2014.
\newblock arXiv:1402.3108 [quant-ph].

\bibitem{experiment_realizing_2022}
Sebastian Krinner and Lacroix et~al.
\newblock Realizing {Repeated} {Quantum} {Error} {Correction} in a {Distance}-{Three} {Surface} {Code}.
\newblock {\em Nature}, 605(7911):669--674, May 2022.
\newblock arXiv:2112.03708 [cond-mat, physics:quant-ph].

\bibitem{experiment_correlated_2021}
C.~D. Wilen, S.~Abdullah, N.~A. Kurinsky, C.~Stanford, L.~Cardani, G.~D'Imperio, C.~Tomei, L.~Faoro, L.~B. Ioffe, C.~H. Liu, A.~Opremcak, B.~G. Christensen, J.~L. DuBois, and R.~McDermott.
\newblock Correlated {Charge} {Noise} and {Relaxation} {Errors} in {Superconducting} {Qubits}.
\newblock {\em Nature}, 594(7863):369--373, June 2021.
\newblock arXiv:2012.06029 [cond-mat, physics:quant-ph].

\bibitem{experiment_mid-circuit_2022}
Kevin Singh, Conor~E. Bradley, Shraddha Anand, Vikram Ramesh, Ryan White, and Hannes Bernien.
\newblock Mid-circuit correction of correlated phase errors using an array of spectator qubits, August 2022.
\newblock arXiv:2208.11716.

\bibitem{experiment1}
Rajeev Acharya and Aghababaie-Beni et~al.
\newblock Quantum error correction below the surface code threshold, August 2024.
\newblock arXiv:2408.13687.

\bibitem{experiment2}
Shingo Kono, Jiahe Pan, Mahdi Chegnizadeh, Xuxin Wang, Amir Youssefi, Marco Scigliuzzo, and Tobias~J. Kippenberg.
\newblock Mechanically {Induced} {Correlated} {Errors} on {Superconducting} {Qubits} with {Relaxation} {Times} {Exceeding} 0.4 {Milliseconds}, May 2023.
\newblock arXiv:2305.02591 [quant-ph].

\bibitem{correlated_fowler2014quantifying}
Austin~G Fowler and John~M Martinis.
\newblock Quantifying the effects of local many-qubit errors and nonlocal two-qubit errors on the surface code.
\newblock {\em Physical Review A}, 89(3):032316, 2014.

\bibitem{correlated_quantum_2019}
Naomi~H. Nickerson and Benjamin~J. Brown.
\newblock Analysing correlated noise on the surface code using adaptive decoding algorithms.
\newblock {\em Quantum}, 3:131, April 2019.
\newblock arXiv:1712.00502 [quant-ph].

\bibitem{correlatedthreshold_impact_2021}
B.~D. Clader, Colin~J. Trout, Jeff~P. Barnes, Kevin Schultz, Gregory Quiroz, and Paraj Titum.
\newblock Impact of correlations and heavy-tails on quantum error correction.
\newblock {\em Physical Review A}, 103(5):052428, May 2021.
\newblock arXiv:2101.11631 [quant-ph].

\bibitem{kitaev_surfacecode}
A~Yu Kitaev.
\newblock Fault-tolerant quantum computation by anyons.
\newblock {\em Annals of physics}, 303(1):2--30, 2003.

\bibitem{correlated_2023learning}
Robin Harper and Steven~T Flammia.
\newblock Learning correlated noise in a 39-qubit quantum processor.
\newblock {\em PRX Quantum}, 4(4):040311, 2023.

\bibitem{correlated_tiurev2023correcting}
Konstantin Tiurev, Peter-Jan~HS Derks, Joschka Roffe, Jens Eisert, and Jan-Michael Reiner.
\newblock Correcting non-independent and non-identically distributed errors with surface codes.
\newblock {\em Quantum}, 7:1123, 2023.

\bibitem{NNNcrosstalk_marxer2023long}
Fabian Marxer, Antti Veps{\"a}l{\"a}inen, Shan~W Jolin, Jani Tuorila, Alessandro Landra, Caspar Ockeloen-Korppi, Wei Liu, Olli Ahonen, Adrian Auer, Lucien Belzane, et~al.
\newblock Long-distance transmon coupler with cz-gate fidelity above 99.8\%.
\newblock {\em PRX Quantum}, 4(1):010314, 2023.

\bibitem{decoder1}
David~K. Tuckett, Stephen~D. Bartlett, Steven~T. Flammia, and Benjamin~J. Brown.
\newblock Fault-tolerant thresholds for the surface code in excess of 5\% under biased noise.
\newblock {\em Physical Review Letters}, 124(13):130501, March 2020.
\newblock arXiv:1907.02554 [cond-mat, physics:quant-ph].

\bibitem{decoder2}
David~K. Tuckett, Andrew~S. Darmawan, Christopher~T. Chubb, Sergey Bravyi, Stephen~D. Bartlett, and Steven~T. Flammia.
\newblock Tailoring {Surface} {Codes} for {Highly} {Biased} {Noise}.
\newblock {\em Physical Review X}, 9(4):041031, November 2019.

\bibitem{surface_code_nishimori}
A~Honecker, M~Picco, and P~Pujol.
\newblock Nishimori point in the 2d+/-j random-bond ising model.
\newblock {\em arXiv preprint cond-mat/0010143}, 2000.

\bibitem{gidney2021stim}
Craig Gidney.
\newblock Stim: a fast stabilizer circuit simulator.
\newblock {\em Quantum}, 5:497, 2021.

\bibitem{pymatching2}
Oscar Higgott and Craig Gidney.
\newblock Sparse blossom: correcting a million errors per core second with minimum-weight matching.
\newblock {\em Quantum}, 9:1600, 2025.

\bibitem{higgott2022pymatching}
Oscar Higgott.
\newblock Pymatching: A python package for decoding quantum codes with minimum-weight perfect matching.
\newblock {\em ACM Transactions on Quantum Computing}, 3(3):1--16, 2022.

\bibitem{circuit_2023improved}
Oscar Higgott, Thomas~C Bohdanowicz, Aleksander Kubica, Steven~T Flammia, and Earl~T Campbell.
\newblock Improved decoding of circuit noise and fragile boundaries of tailored surface codes.
\newblock {\em Physical Review X}, 13(3):031007, 2023.

\bibitem{circuit_chamberland2022universal}
Christopher Chamberland and Earl~T Campbell.
\newblock Universal quantum computing with twist-free and temporally encoded lattice surgery.
\newblock {\em PRX Quantum}, 3(1):010331, 2022.

\bibitem{detuning_kelly2014optimal}
Julian Kelly, Rami Barends, Brooks Campbell, Yu~Chen, Zijun Chen, Ben Chiaro, Andrew Dunsworth, Austin~G Fowler, I-C Hoi, Evan Jeffrey, et~al.
\newblock Optimal quantum control using randomized benchmarking.
\newblock {\em Physical review letters}, 112(24):240504, 2014.

\bibitem{detuning_zhao2022quantum}
Peng Zhao, Kehuan Linghu, Zhiyuan Li, Peng Xu, Ruixia Wang, Guangming Xue, Yirong Jin, and Haifeng Yu.
\newblock Quantum crosstalk analysis for simultaneous gate operations on superconducting qubits.
\newblock {\em PRX quantum}, 3(2):020301, 2022.

\end{thebibliography}
\end{document}